\documentclass{elsart}

\usepackage{amssymb}

\usepackage[final]{graphicx}

\begin{document}

\begin{frontmatter}

\title{Combination of carbon nanotubes and two-photon absorbers for broadband
optical limiting}
\author[DGA,GDPC]{N.Izard}
\author[CEA]{C.M\'enard}
\author[DGA]{D.Riehl\corauthref{cor1}}
\ead{didier.riehl@dga.defense.gouv.fr}
\author[CEA]{E.Doris}
\author[CEA]{C. Mioskowski}
\author[GDPC]{E.Anglaret\corauthref{cor1}}
\ead{eric@gdpc.univ-montp2.fr}
\corauth[cor1]{Corresponding
authors.}
\address[DGA]{Centre Technique d'Arcueil, DGA, Arcueil, France}
\address[GDPC]{Groupe de Dynamique des Phases Condens\'ees, UMR CNRS 5581, Universit\'e Montpellier II, Montpellier,
France}
\address[CEA]{Service de Marquage Mol\'eculaire et de Chimie Bioorganique,CEA-Saclay, Gif sur Yvette, France}

\begin{abstract}
New systems are required for optical limiting against broadband
laser pulses. We demonstrate that the association of non-linear
scattering from single-wall carbon nanotubes (SWNT) and
multiphoton absorption (MPA) from organic chromophores is a
promising approach to extend performances of optical limiters over
broad spectral and temporal ranges. Such composites display high
linear transmission and good neutral colorimetry and are
particularly efficient in the nanosecond regime due to cumulative
effects.
\end{abstract}
\end{frontmatter}

\section{Introduction}
Active eye and sensor protection against high power laser is an
operational need of growing importance due to the democratization
of frequency-agile pulsed lasers. An ideal optical limiter should
fulfill high-level specifications, like broadband optical limiting
efficiency over the whole visible spectrum, and broadband temporal
efficiency from sub-nanosecond pulses to continuous regime.
Furthermore, it should also preserve the quality of
observation/detection at low fluences and should therefore display
high linear optical transmittance and neutral colorimetry.
Extensive researchs were conducted over the past fifteen years and
led to the emergence of three main classes of nonlinear optical
materials : reverse saturable absorbers \cite{RSA}, multiphoton
absorbers \cite{MPA} and nonlinear scattering systems
\cite{CBS,NT}. However, none of these systems, \emph{taken
individually}, is able to fully fulfill the specifications listed
above. Some attempts were performed with combinations of nonlinear
optical materials in cascaded geometries : multi-plate or tandem
cells \cite{cascade}, use of two intermediate focal planes of a
sighting system \cite{twofocal}.

In this letter, we propose a new approach for the design of
composite optical limiters, in which two complementary nonlinear
optical materials are mixed together in the same cell. Indeed, we
studied mixtures of single wall carbon nanotubes (SWNT)
suspensions, which were recently shown to be effective optical
limiters  \cite{NT}, and multiphoton absorbers (MPA) solutions.
Nonlinear scattering is the main limiting phenomenon for SWNT
suspensions, due to the growth of solvent vapor bubbles at low
fluences and long pulses, and to the growth of carbon bubbles from
nanotube sublimation at high fluences and short pulses
\cite{vivien}. Optical limiting is effective for relatively long
pulses, typically from the nanosecond to the microsecond regimes.
As far as MPA solutions are concerned, their strong multiphoton
absorption cross-sections are responsible of limitation, which is
effective for shorter pulses (subpicosecond to a few nanosecond).
The materials and the experimental aspects of the work are
presented in section 2. The association of the non linear
scattering properties of SWNT and the multiphoton absorption
properties of MPA is investigated in the nanosecond regime in
section 3. We finally discuss the expected optical limiting
performances of such composites from picosecond to microsecond
regimes in section 4.

\section{Materials and Experimental}
SWNT were produced by the electric arc process and were purchased
from Nanoledge$^{\circledR}$ and MER$^{\circledR}$. These samples
were extensively characterised using scanning and transmission
electron microscopy, X-Ray diffraction, Raman and optical
spectroscopy \cite{Kirchberg_Nico}. The mean diameter of the
nanotubes is approximately of 1.4 nm. Suspensions of as-produced,
raw samples (from Nanoledge$^{\circledR}$), hereafter referred as
NT-Raw, were prepared in chloroform \cite{NT}. By contrast,
purified nanotubes (from MER$^{\circledR}$) could not be suspended
in chloroform. These purified nanotubes were made soluble in
chloroform by grafting long alkyl chains on their surface
\cite{Riggs}. Indeed, acidic treatment (H$_{2}$SO$_{4}$ /
HNO$_{3}$ (3/1)) of the nanotubes leads to the formation of
carboxylic groups on the nanotube surface \cite{fonctions}.
Activation of these groups with thionyl chloride followed by
coupling with octadecylamine yields grafted nanotubes which are
fully soluble in chloroform \cite{greffage}. The linear
transmittance spectra of suspensions of raw and functionalised
samples (hereafter designated as NT-Graft) are almost
undistinguishable (not shown). The optical limiting properties of
as-produced nanotubes and grafted purified nanotubes suspensions
in chloroform are similar, whatever the pulse duration.

2,2'-([1,1'-biphenyl]-4,4'-diyldi-2,1-ethenediyl)-bis-benzenesulfonic
acid disodium salt, most usually named \textit{Stilbene-3}
\cite{stilbene}, is a commercially available dye which was used as
reference compound for multi-photon absorption. Solubility of
stilbene-3 is high (300 g/l) in DMSO and moderate (20 g/l) in
water but stilbene-3 is insoluble in chloroform. In order to
achieve solubility of stilbene-3 in chloroform, sodium
counter-ions were exchanged with quaternary dimethyl-dioctadecyl
ammonium groups. The presence of ammoniums bearing long alkyl
chains on the dye allowed modified stilbene-3 to be soluble in
CHCl$_{3}$ ($>$200 g/l).

Linear optical transmission of SWNT suspensions were adjusted to
70 $\%$ at 532 nm in 2 mm thick cells. For such transmittances,
the nanotube concentration was around 10 mg/l. The MPA
concentration in the reference sample and in the composites was 50
g/l. Nonlinear optical transmittance measurements were performed
using different laser sources and test beds : (i) a RSG-19
experimental set-up in a F/5 focusing geometry with a 1.5 mrad
collecting aperture \cite{otan}, using a frequency-doubled Nd:YAG
laser emitting 7 ns pulses at 532 nm (ii) an optical parametric
oscillator (OPO) with a pulse duration of 3 ns at 532 nm, in a
F/30 focusing geometry, (iii) a Q-switched, but non injected,
frequency doubled Nd:YAG, with a pulse duration of 15 ns in a F/50
focusing geometry. Note that for the two latter test beds, the
Rayleigh length is larger than the cell length and no diaphragm is
used. Therefore, results from these two test beds and RSG-19 set
up can not be directly compared.

\section{Results}
\begin{figure}
\begin{center}
\includegraphics[width=15cm]{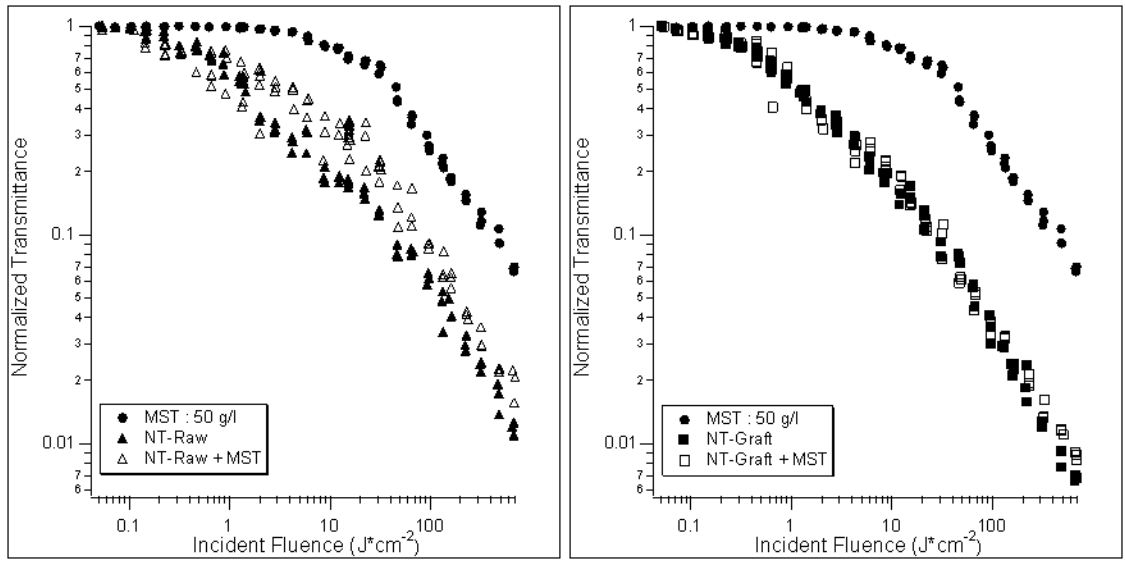}
\caption{Optical limiting results at 532 nm, for 7 ns pulses, F/5
geometry and small aperture. The results for the reference samples
(SWNT and modified stilbene-3) are compared to those of the
mixtures for raw nanotube samples (left) and functionalised
nanotubes (right).} \label{fig:Fig_OTAN}
\end{center}
\end{figure}

The results obtained in the F/5 geometry, where the focusing
conditions allow to work at high fluences, are reported in figure
\ref{fig:Fig_OTAN}.

Unexpectedly, the optical limiting performances are worse for the
combination of nanotubes and modified stilbene-3 than for the
nanotubes alone (figure \ref{fig:Fig_OTAN}, left). In the mixture,
limiting is thwarted by an adverse effect. The drop of
performances is especially important at low fluences. We also
observed that the stability of the composite suspension was poor :
aggregation of the nanotubes occurred within a few hours. By
contrast, when the suspensions of SWNT and stilbene-3 were
prepared in water with the help of surfactants \cite{sandiego}, no
instability was observed and the optical limiting performances of
the mixture were slightly better than those of the nanotubes
alone. On the other hand, if grafted SWNT are used (figure
\ref{fig:Fig_OTAN}, right), the adverse effect disappears although
no improvement is observed.

\begin{figure}
\begin{center}
\includegraphics[width=15cm]{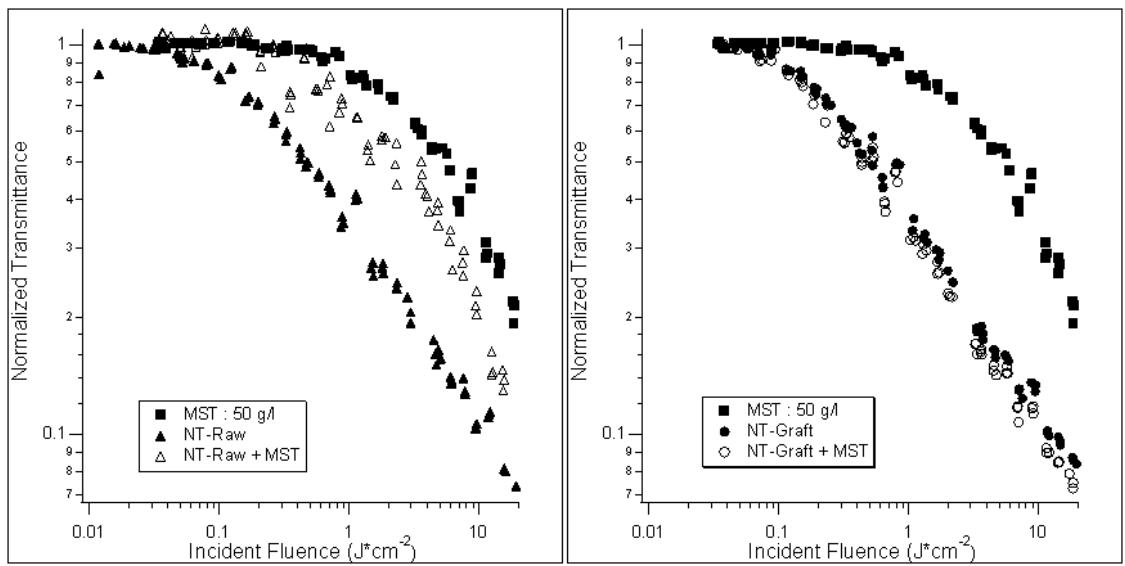}
\caption{Optical limiting results at 532 nm, for 15 ns pulses, F/50
geometry. The results for the reference samples
are compared to those of the mixtures for raw
nanotube samples (left) and functionalised nanotubes (right).}
\label{fig:Fig_NICO}
\end{center}
\end{figure}

The same samples were studied with a different geometry (F/50) and
longer pulses (15 ns). Results are displayed in figure
\ref{fig:Fig_NICO}. The loss of performances in the composite prepared with raw
nanotubes is striking (fig. \ref{fig:Fig_NICO}, left). The optical
limiting threshold is significantly larger than that of the
nanotubes alone (200 mJ/cm$^{-2}$ for the mixture vs 50
mJ/cm$^{-2}$ for raw nanotubes). Once again, no adverse effect is
observed with composites prepared with functionalised nanotubes.
Furthermore, one observes a slight but significant cumulative
effect, as initially expected (fig. \ref{fig:Fig_NICO}, right).

In the light of these experiments, we assign the adverse effect to
an adsorption of the chromophores on the surface of the nanotubes.
Such a coating acts as a contact resistance which delays heating of
the surrounding solvent. A similar effect has been observed on
carbon black suspensions embedded in surfactant \cite{pistack}.
The hypothesis of contact resistance is also confirmed by
pump-probe experiments which will be reported elsewhere
\cite{nico}. Adsorption may be due to $\pi$-stacking of the
aromatic rings of the MPA over the nanotube surface. In water
suspensions, the presence of surfactants prevents the formation of
the coating \cite{sandiego}. For functionalised samples,
adsorption is hindered by the octadecylamine chains grafted on the
nanotube surface which act as steric barriers.

\begin{figure}
\begin{center}
\includegraphics[width=15cm]{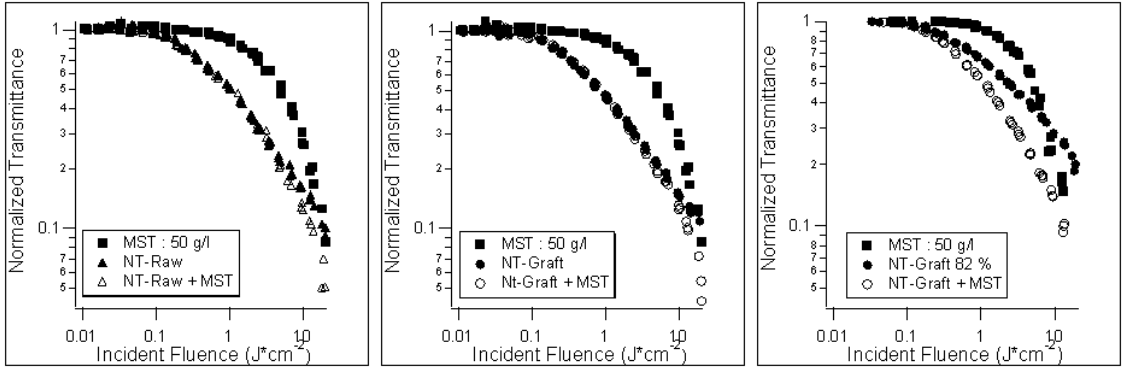}
\caption{Optical limiting results at 532 nm, for 3 ns pulses, F/30
geometry. The results for the reference samples are compared to
those of the mixtures for raw nanotube samples (left) and for
functionalised nanotubes, at regular and low concentrations (70
and 82$\%$ transmittance) in the middle and right graphs,
respectively.} \label{fig:Fig_OPO_532}
\end{center}
\end{figure}

Finally, optical limiting experiments were performed with 3 ns
pulses. Results are shown in figure \ref{fig:Fig_OPO_532}. With
this set-up, the results are very similar for raw and grafted
nanotubes. This is in agreement with the hypothesis of a contact
resistance between chromophores and nanotubes. Indeed, for short
(3 ns) pulses, nonlinear scattering is mainly due to the
sublimation of nanotubes which induces the growth of carbon vapor
bubbles \cite{vivien}. By contrast, for longer (15 ns) pulses,
scattering at the limiting threshold is due to the growth of
solvent vapor bubbles, nucleated by heat transfert from SWNT to
surrounding solvent \cite{vivien}. While coating of the
chromophore on the nanotubes is expected to slow down the heat
transfer from SWNT to solvent, one does not expect any change in
the heating and sublimation dynamics of the nanotubes.

\begin{figure}
\begin{center}
\includegraphics[width=15cm]{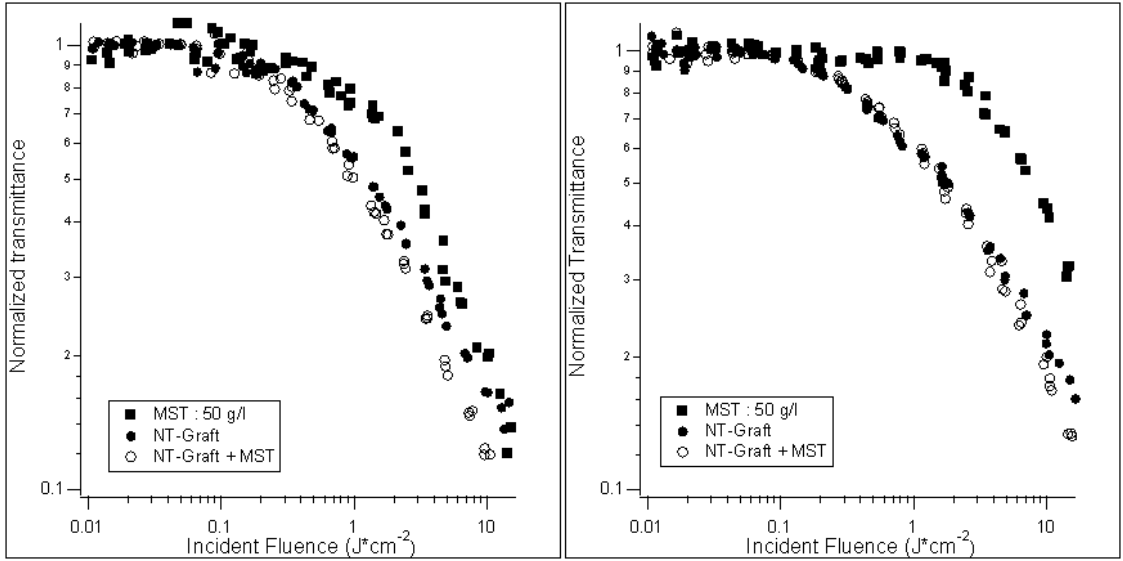}
\caption{Optical limiting curves for 3 ns pulses, F/30 geometry at
450 nm (left) and 650 nm (right).} \label{fig:Fig_broadband}
\end{center}
\end{figure}

The grafting of long alkyl chains on the nanotubes prevent
interactions with the chromophores and allow cumulated
efficiencies of both SWNT and MPA. The improvement of the
performances due to the cumulative effect is best observed when
the limitation efficiency of the chromophores is comparable to
that of the nanotubes. This occurs only at high fluences for
nanosecond pulses and 70 $\%$ transmittance cells. In the right
part of figure \ref{fig:Fig_OPO_532}, the cumulative effect is
emphasized for a sample prepared with a three times smaller
concentration of nanotubes (82 $\%$ transmittance). It is obvious
that the cumulative effect will also be more effective for larger
concentrations of chromophore and/or chromophores with larger
multiphoton absorption cross-sections \cite{sandiego}. On the
other hand, both non linear scattering from nanotubes and
multiphoton absorption are expected to be effective over a broad
spectral range. This is confirmed by measurements at various
wavelengths, which are displayed in figure
\ref{fig:Fig_broadband}. The best performances are obtained at
small wavelengths, which are associated to nanotubes larger
absorbing cross-sections and vapor bubbles larger scattering
cross-sections according to Mie theory.

\section{Conclusion}
The concept of optical limiting systems based on the mixture of
carbon nanotubes and multiphoton absorber chromophores was
demonstrated. Stable composite suspensions can be prepared, using
nanotubes grafted with long alkyl chains. This strategy also
inhibits coating of the chromophores on the nanotube surface. We
studied the optical limiting properties of the mixtures in the ns
range. When the limiting efficiencies of the two components are
close, a cumulative effect is observed. This occurs especially for
3 ns pulses in the case of SWNT/MST mixtures. The range of pulse
durations where the cumulative effect will be effective can be
easily broadened by increasing the chromophore concentration
and/or selecting chromophores with larger multiphoton absorption
cross-sections. When the limiting performances of one of the
component dominate those of the other, the limiting properties are
close to those of the one-component system. This is what we
observe for pulses of 7 or 15 ns in the case of SWNT/MST mixtures.
Note that optical limiting is effective for suspensions of
nanotubes up to microsecond pulses. On the other hand, MPA are
good limiters in the sub-nanosecond range. Both nonlinear
scattering by nanotubes and non linear absorption by MPA are
effective all over the visible range. Therefore, composites
SWNT/MPA are promising systems for optical limiting on broad
temporal and spectral ranges.

\textbf{Acknowledgments}

This work benefited from fruitful interactions with J. Delaire, F.
Lafonta, C. Andraud, M. Blanchard-Desce.



\begin{thebibliography}{00}

\bibitem{RSA}
J.S.Shirk, R.G.S.Pong, S.R.Elom, M.E.Boyle, A.W.Snow, MRS
Proceedings 374 (1995) 201-209.

\bibitem{MPA}
J.W.Perry, S.Barlow, J.E.Ehrlich, A.A.Heikal, Z.-Y.Hu, I.-Y.S.Lee,
K.Mansour, S.R.Marder, H.R\"\o ckel, M.Rumi, S.Thayumanavan,
X.L.Wu, Nonlinear Optics 21 (1999) 225-243.

\bibitem{CBS}
K.J.McEwan, P.K.Milsom, D.B.James, SPIE Proceedings 3472 (1998)
47-52.

\bibitem{NT}
L.Vivien, E.Anglaret, D.Riehl, F.Bacou, C.Journet, C.Goze,
M.Andrieux, M.Brunet, F.Lafonta, P.Bernier, F.Hache, Chemical
Physics Letters 307 (1999) 317 and erratum, \textit{ibid} 312
(2000) 617.

\bibitem{vivien}
L.Vivien, P.Lan\c con, D.Riehl, F.Hache, E.Anglaret, Carbon 40
(2002) 1789.

\bibitem{cascade}
P.A.Miles, Applied Optics 33 (1994) 6965.

\bibitem{twofocal}
E.W.Van Stryland, S.S.Yang, F.E.Hernandez, V.Dubikovsky,
W.Shensky, D.J.Hagan, Nonlinear Optics 27 (2001) 181.

\bibitem{Kirchberg_Nico}
N.Izard, D.Riehl, E.Anglaret, AIP Conference Proccedings Vol 685,
Molecular Nanostructures, (2003) 235-240.

\bibitem{Riggs}
J.E.Riggs, D.B.Walker, D.L.Caroll, Y.-P.Sun, Journal of Physical
Chemistry B 30 (2000) 7071-7076.

\bibitem{fonctions}
J.Liu, A.G.Rinzler, H.Dai, J.H.Hafner, R.K.Bradley, P.J.Boul,
A.Lu, T.Iverson, K.Shelimov, C.B.Huffman, F.Rodriguez-Macias,
Y-S.Shon, T.T.Lee, D.T.Colbert, R.E.Smalley, Science 280 (1998)
1253.

\bibitem{greffage}
J.Chen, M.A.Hamon, H.Hu, Y.Chen, A.M.Rao, P.C.Eklund, R.C.Haddon,
Science 282 (1998) 95.

\bibitem{stilbene}
P.-A.Chollet, V.Dumarcher, J.-M. Nunzi, P.Feneyrou, P.Baldek,
Nonlinear Optics 21 (1999) 299-308.

\bibitem{otan}
D.B.James, K.J.McEwan, Nonlinear Optics 21 (1999) 377-389.

\bibitem{sandiego}
D.Riehl, N.Izard, L.Vivien, E.Anglaret, E.Doris, C.M\'enard,
C.Mioskowski, L. Porr\`es, O.Mongin, M.Charlot, M.Blanchard-Desce,
R.An\'emian, J.-C.Mulatier, C.Barsu, C.Andraud, Proceedings of
SPIE Vol 5211, Nonlinear Optical Transmission and Multiphoton
Processes in Organics, (2003) 124-134.

\bibitem{pistack}
F.Fougeanet, D.Riehl, Nonlinear Optics 21 (1999) 435-446.

\bibitem{nico}
N.Izard \textit{et al.}, unpublished results.

\end{thebibliography}
\end{document}